# On the key kinetic interactions between NOx and unsaturated hydrocarbons: H-atom abstraction from $C_3$-$C_7$ alkynes and dienes by $NO_2$

Zhengyan Guo[a,d,^], Hongqing Wu[b,^], Ruoyue Tang[b,^], Xinrui Ren[b], Ting Zhang[b], Mingrui Wang[b], Guojie Liang[b], Hengjie Guo[a], Song CHENG[b,c,]*

[a] *School of Power and Energy, Northwestern Polytechnical University, Xi'an 710129, China*
[b] *Department of Mechanical Engineering, The Hong Kong Polytechnic University, Kowloon, Hong Kong SAR, China*
[c] *Research Institute for Smart Energy, The Hong Kong Polytechnic University, Kowloon, Hong Kong SAR, China*
[d] *AECC Hunan Aviation Powerplant Research Institute, Zhuzhou 412002, China*

^ Authors contribute equally to this paper.
* Corresponding authors: Song Cheng
Phone: +852 2766 6668
Email: songryan.cheng@polyu.edu.hk

**Abstract:**

An adequate understanding of $NO_x$ interacting chemistry is a prerequisite for a smoother transition to carbon-lean and carbon-free fuels such as ammonia and hydrogen. In this regard, this study presents a comprehensive study on the H-atom abstraction by $NO_2$ from $C_3$-$C_7$ alkynes and dienes forming 3 $HNO_2$ isomers (i.e., TRANS_HONO, HNO2, and CIS_HONO), encompassing 8 hydrocarbons and 24 reactions. Through a combination of high-level quantum chemistry computation, electronic structures, single point energies, C-H bond dissociation energies and 1-D hindered rotor potentials of the reactants, transition states, complexes and products involved in each reaction are determined at DLPNO-CCSD(T)/cc-pVDZ//M06–2X/6−311++g(d,p), from which potential energy surfaces and energy barriers for each reaction are determined. Following this, the rate coefficients for all studied reactions, over a temperature range from 298 to 2000 K, are computed based on Transition State Theory using the Master Equation System Solver program with considering unsymmetric tunneling corrections. Comprehensive analysis of branching ratios elucidates the diversity and similarities between different species, different $HNO_2$ isomers, and different abstraction sites. Incorporating the calculated rate parameters into a recent chemistry model reveals the significant influences of this type of reaction on model performance, where the updated model is consistently more reactive for all the alkynes and dienes studied in predicting autoignition characteristics. Sensitivity and flux analyses are further conducted, through which the importance of H-atom abstractions by $NO_2$ is highlighted. With the updated rate parameters, the branching ratios in fuel consumption clearly shifts towards H-atom abstractions by $NO_2$ while away from H-atom abstractions by ȮH. The obtained results emphasize the need for adequately representing these kinetics in new alkyne and diene chemistry models to be developed by using the rate parameters determined in this study, and call for future efforts to experimentally investigate $NO_2$ blending effects on alkynes and dienes.

# 1. Introduction

The need for sufficient understanding of $NO_x$/hydrocarbon interactions has been more imminent than ever. On one hand, the recently growing interests in using ammonia as carbon-free alternative fuel for fossil-derived conventional fuels have advocated fuel-blending combustion as a means to address the poor combustion performance and high $NO_x$ emissions of pure ammonia. Under this strategy, ammonia is typically bended with a more reactive hydrocarbon (e.g., n-heptane [1]), where the $NO_x$ and $NH_x$ species produced from ammonia chemistry can alter substantially the chemistry of the companion fuel. On the other hand, from the perspective of advanced engine operation, exhaust gas recirculation (EGR) has a long history in application to advanced internal combustion engines, enabling better and wider control of combustion phasing/heat release rates (HRRs) in homogeneous charge compression ignition engines [2] and knock in spark-ignition engines [3]. The use of EGR introduces both physical and chemical effects on engine combustion performance, among which the chemical influences of minor EGR species such as $NO_x$ have been found to be significant [4] even at small in-cylinder concentrations (e.g., 10 – 250 ppm).

As such, in recent decades, extensive fundamental studies have been carried out for investigating the interactions between $NO_x$ and typical hydrocarbons [4-9]. Both inhibiting and promoting effects of $NO_x$ on fuel reactivity have been reported, highlighting the complicated $NO_x$/hydrocarbon interacting chemistries. While some understandings of the underlying kinetics have been gained through these studies, the existing studies still fall short in the following: (a) past studies mainly focused on NO with $NO_2$ largely overlooked; and (b) there is a severe lack of studies on unsaturated hydrocarbons, with most of the existing studies focusing on alkanes and aromatics. Menon et al. [10] examined the influence of $NO_2$ on the formation of soot precursors during the pyrolysis of ethylene in a flow reactor and reported significant impact of $NO_2$ on ethylene consumption and soot precursor formation. Deng et al.

[11,12] investigated high-pressure ignition characteristics of $NO_2/C_2H_4$ mixtures and $NO_2/C_3H_6$ mixtures in a shock tube. The experiments illustrated a significant decrease in the ignition delay time of ethylene and propylene with the presence of $NO_2$. Yuan et al. [13] conducted experimental and modeling analysis to explore the interaction kinetics between $NO_2$ and propylene in a laminar flow reactor. They found that the direct interactions between $NO_2$ and the fuel molecules and their primary derivatives are the major causes for the changed reactivity. Very recently, Cheng et al. [14] characterized the interactions between $NO_x$ and ethylene/propylene/isobutylene in a rapid compression machine, and revealed the strong influences of these interactions on the autoignition reactivity of gasoline fuels. Through limited, these studies have consistently revealed a major type of interaction reactions directly involving $NO_2$, namely $RH+NO_2=\dot{R}+HNO_2/HONO$, that greatly promotes reactivity.

There have been a few experimental and theoretical studies to determine the rate coefficients of $RH+NO_2=\dot{R}+HNO_2/HONO$. Chai et al. [15] calculated the rate coefficients for H-atom abstraction from several alkanes and alkenes by $NO_2$ at the CCSD(T)-F12a/cc-pVTZ-f12//B2PLYPD3/cc-pVTZ level of theory. Wang et al. [16] calculated the H-atom abstraction from n-decane by $NO_2$ at CBS-QB3//M06–2X/6–311++G (d,p). More recently, Wu et al. [17] determine the rate coefficients for H-atom abstractions by $NO_2$ from $C_2$-$C_5$ alkanes and alkenes at the DLPNO-CCSD(T)/cc-pVDZ//M06–2X/6−311++g(d,p) level of theory. Nevertheless, there are no theoretical studies on H-atom abstractions by $NO_2$ from alkynes and dienes. As a result, these reactions are typically missing in existing chemistry models or the rate coefficients are defined by analogy from alkanes and alkenes, leading to large uncertainties in model predictions. Although Ohta et al. [18] measured the rate coefficients of reactions between $NO_2$ and 16 conjugated alkanes. The reported rate coefficients were combined rate coefficients without information on reaction products or reaction site.

Therefore, the aims of this study are threefold: (a) conducting a detailed theoretical investigation of H-atom abstractions from $C_3$-$C_7$ alkynes and dienes by $NO_2$ that form $HNO_2$, TRANS_HONO and CIS_HONO, via quantum chemistry computation and statistical rate theories; (b) reveal the branching ratios of the three pathways forming the three HNO2 isomers for the selected species at different H-atom sites and different size of molecules; and (c) systematically analyze the effects of these reactions on model performance and the underlying kinetics.

## 2. Computational methods

2.1. Potential energy surfaces

Electronic geometries, vibrational frequencies and zero-point energies for all species involved in the 24 reactions (including reactants, products, complexes, transition states (TS)) are calculated at the M06-2X method [19] coupled with the 6-311++G(d,p) basis set [20-22]. Conformer search at the same level of theory is conducted to ensure the optimized structures retain the lowest energy. Intrinsic reaction coordinate (IRC) calculations have been carried out at the same level of theory to ensure that the transition state connects the respective reactants with the respective product complex. 1-D hindered rotor treatment [23] are also obtained at the M06–2X/6–311++G(d, p) level of theory for the low frequency torsional modes between non-hydrogen atoms in all of the reactants, TS, complexes and products, with a total of 18 scans (i.e., 20 degrees increment in the respective dihedral angle) for each rotor. Scale factors of 0.983 for harmonic frequencies and 0.9698 for ZPEs that were recommended by Zhao and Truhlar [19] are used herein. Single-point energies (SPEs) are further determined for all the species using the DLPNO-CCSD(T) functional [24-25] with the cc-pVDZ basis set. With the CCSD(T) method, attention must be addressed to T1 diagnostic [26] to measure the multi-reference effect. The T1 diagnostic values for all the species, as summarized in Table S1 in the Supplementary Material, are below 0.029, which indicates that the SPEs calculated from using single-reference calculation method are reliable in this study. All the calculations mentioned above are performed using ORCA 5.0.4 [27], and the optimized structures for all species, TS and complexes are summarized in the Supplementary Material.

2.2. Rate constant calculations

The Master Equation System Solver (MESS) program suite [28] is employed here to calculate the chemical rate coefficients for complex-forming reactions via solution of the one-dimensional master equation, based on the chemically significant eigenstate approach of Miller and Klippenstein [29] and the bimolecular species model of Georgievskii and Klippenstein [30]. The frequencies of lower-frequency modes are replaced by the hindered rotor potentials obtained from 1-D scans. Quantum mechanical tunneling corrections assuming the asymmetric Eckart potential (TST/Eck) [31] are applied to obtain the rate coefficient over the temperature range of 298–2000 K. All rate coefficients were fitted to the modified Arrhenius equation, which can be defined as $k = AT^n \exp(-E_a/RT)$.

2.3. Kinetic modeling

The latest chemistry model developed by LLNL [32] is used herein to investigate the influence caused by these calculated reactions on the model performance in predicting combustion characteristics of alkynes and dienes. The development of the chemistry model has been documented in [32], hence will not be presented here. To better represent the benchmark kinetics, the rate coefficients of the H-atom abstraction reactions by $CH_3\dot{O}$ and $CH_3O\dot{O}$ radicals have also been updated following recent studies [33-34]. Kinetic modeling of fundamental combustion experiments, namely autoignition experiments, are completed using the LLNL-developed fast solver Zero-RK [35].

# 3. Results and discussion

## 3.1. Species and reaction sites

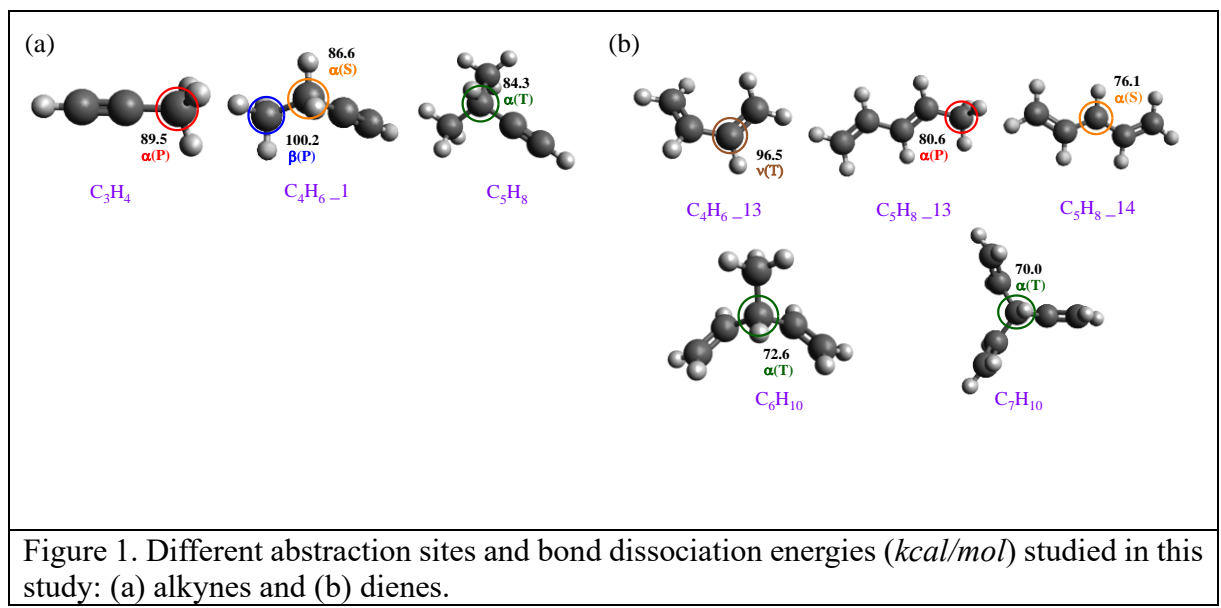

Figure 1. Different abstraction sites and bond dissociation energies (*kcal/mol*) studied in this study: (a) alkynes and (b) dienes.

This study involves three alkynes ($C_3H_4$, $C_4H_6\_1$, and $C_5H_8$) and five dienes ($C_4H_6\_13$, $C_5H_8\_13$, $C_5H_8\_14$, $C_6H_{10}$, and $C_7H_{10}$). Each species can undergo H-atom abstraction by $NO_2$ to form three $HNO_2$ isomers (TRANS_HONO, $HNO_2$, and CIS_HONO), encompassing a total of 24 reactions. To better illustrate these reactions, the reaction sites for each species are marked in Figure 1. According to the type of C-atoms to which H-atoms bond, the reaction sites are divided into primary (P), secondary (S), and tertiary (T) sites. Additionally, reaction sites are designated as α and β according to the proximity to the functional group, while those located at C=C bonds are defined as ν.

Furthermore, C-H bond dissociation energies (BDEs) for different reaction sites at 298 K are also calculated via

$$BDE_{298}(R-H) = \Delta H^0_{f,298}(\dot{R},g) + \Delta H^0_{f,298}(\dot{H},g) - \Delta H^0_{f,298}(RH,g)$$

As can be seen in Fig. 1a, the BDEs at different sites of alkynes rank as follows, from highest to lowest: β(P) > α(P) > α(S) > α(T), with differences between the P, S, and T sites less than 6 kcal/mol. Notably, the BDE of the β(P) site in $C_4H_6\_1$ is significantly higher than that of the α(P) site in $C_3H_4$ by approximately 10.7 kcal/mol, highlighting significant impact of functional groups on BDE. For dienes (as shown in Fig. 1b), the BDE at α-carbons is the highest at the P site (80.6 kcal/mol), followed by the S site (76.1 kcal/mol), and then the T sites (72.6 and 70.0 kcal/mol). Compared to the α-carbon sites, the ν site requires even higher energy to break the C-H bond, with a notable difference of about 16 kcal/mol.

## 3.2. Electronic energy barriers

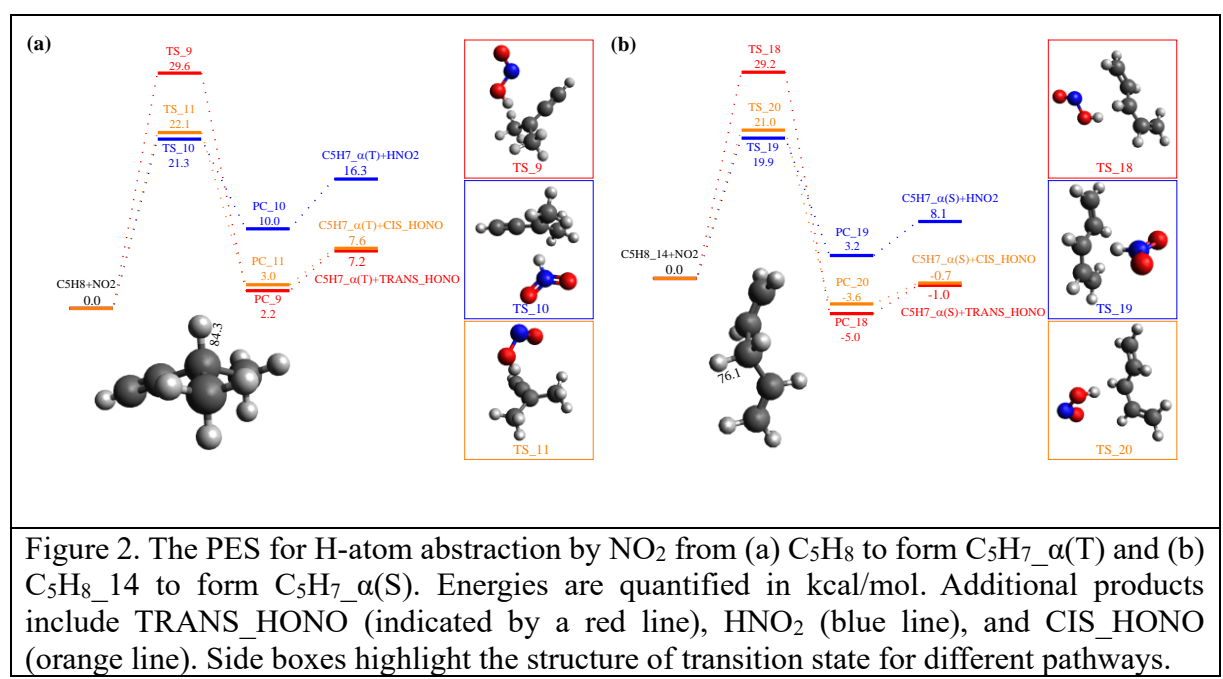

Figure 2. The PES for H-atom abstraction by $NO_2$ from (a) $C_5H_8$ to form $C_5H_7\_α(T)$ and (b) $C_5H_8\_14$ to form $C_5H_7\_α(S)$. Energies are quantified in kcal/mol. Additional products include TRANS_HONO (indicated by a red line), $HNO_2$ (blue line), and CIS_HONO (orange line). Side boxes highlight the structure of transition state for different pathways.

The potential energy surfaces (PES) for H-atom abstractions by $NO_2$ from $C_5H_8$ and $C_5H_8\_14$ are illustrated in Fig. 2a and Fig. 2b, respectively. This figure also includes all related product complexes (PCs) that connect transition states (TS) with the final products. The optimized structures of the transition states are also displayed. Notably, the energy barriers associated with forming TRANS_HONO are consistently the highest among the three channels, which is followed by CIS_HONO and then HNO$_2$. Mebel et al. [36] attributed this phenomenon to the unique interaction between NO$_2$ and the H atom to form TRANS_HONO, during which the ON π bond on NO$_2$ is disrupted, and the unpaired electron is shifted to the oxygen atom to participate in the formation of the O···H bond. This rearrangement increases repulsive interactions between NO$_2$ and the fuel molecule, thereby elevating the energy barrier for TRANS_HONO formation. Similarly, Chai et al. [15] suggest that the high energy barrier for the TRANS_HONO pathway primarily stems from the need for the RH fragment to approach NO$_2$ more closely. This is necessary to achieve sufficient orbital overlap without cancellation, which in turn leads to heightened Coulombic repulsion. The PES for all other species involved in this study exhibit the same trends as those observed in Fig. 2, which are presented in Figs. S1-S2 in the Supplementary Material.

Table 1 The relative energy for H-atom abstraction by NO$_2$ from different sites of alkynes and dienes to form the respective products and HNO$_2$ isomers (TRANS_HONO, HNO$_2$, CIS_HONO). All values are in kcal/mol.

| No. | Reaction | Reactant | TS | Product complex | Product |
|---|---|---|---|---|---|
| | Alkynes + NO$_2$ | | | | |
| R1 | C$_3$H$_4$+NO$_2$→C$_3$H$_3$_α(P)+TRANS_HONO | 0 | 34.4 | 8.4 | 12.4 |
| R2 | C$_3$H$_4$+NO$_2$→C$_3$H$_3$_α(P)+HNO$_2$ | 0 | 27.4 | 17.5 | 21.5 |
| R3 | C$_3$H$_4$+NO$_2$→C$_3$H$_3$_α(P)+CIS_HONO | 0 | 27.4 | 9.0 | 12.8 |
| R4 | C$_4$H$_6$_1+NO$_2$→C$_4$H$_5$_β(P)+TRANS_HONO | 0 | 32.8 | 21.1 | 23.1 |
| R5 | C$_4$H$_6$_1+NO$_2$→C$_4$H$_5$_β(P)+HNO$_2$ | 0 | 31.2 | 27.5 | 32.2 |
| R6 | C$_4$H$_6$_1+NO$_2$→C$_4$H$_5$_β(P)+CIS_HONO | 0 | 29.4 | 20.8 | 23.5 |
| R7 | C$_4$H$_6$_1+NO$_2$→C$_4$H$_5$_α(S)+HNO$_2$ | 0 | 24.0 | 13.3 | 18.6 |
| R8 | C$_4$H$_6$_1+NO$_2$→C$_4$H$_5$_α(S)+CIS_HONO | 0 | 24.0 | 6.2 | 9.8 |
| R9 | C$_5$H$_8$+NO$_2$→C$_5$H$_7$_α(T)+TRANS_HONO | 0 | 29.6 | 2.2 | 7.2 |
| R10 | C$_5$H$_8$+NO$_2$→C$_5$H$_7$_α(T)+HNO$_2$ | 0 | 21.3 | 10.0 | 16.3 |
| R11 | C$_5$H$_8$+NO$_2$→C$_5$H$_7$_α(T)+CIS_HONO | 0 | 22.1 | 3.0 | 7.6 |
| Dienes + NO$_2$ | | | | | |
| R12 | C$_4$H$_6$_13+NO$_2$→C$_4$H$_5$_v(T)+TRANS_HONO | 0 | 35.3 | 15.1 | 19.4 |
| R13 | C$_4$H$_6$_13+NO$_2$→C$_4$H$_5$_v(T)+HNO$_2$ | 0 | 30.7 | 23.9 | 28.5 |
| R14 | C$_4$H$_6$_13+NO$_2$→C$_4$H$_5$_v(T)+CIS_HONO | 0 | 29.7 | 16.0 | 19.7 |
| R15 | C$_5$H$_8$_13+NO$_2$→C$_5$H$_7$_α(P)+TRANS_HONO | 0 | 29.4 | -0.6 | 3.5 |
| R16 | C$_5$H$_8$_13+NO$_2$→C$_5$H$_7$_α(P)+HNO$_2$ | 0 | 22.3 | 8.2 | 12.6 |
| R17 | C$_5$H$_8$_13+NO$_2$→C$_5$H$_7$_α(P)+CIS_HONO | 0 | 23.5 | 0.5 | 3.9 |
| R18 | C$_5$H$_8$_14+NO$_2$→C$_5$H$_7$_α(S)+TRANS_HONO | 0 | 29.2 | -5.0 | -1.0 |
| R19 | C$_5$H$_8$_14+NO$_2$→C$_5$H$_7$_α(S)+HNO$_2$ | 0 | 19.9 | 3.2 | 8.1 |
| R20 | C$_5$H$_8$_14+NO$_2$→C$_5$H$_7$_α(S)+CIS_HONO | 0 | 21.0 | -3.6 | -0.7 |
| R21 | C$_6$H$_{10}$+NO$_2$→C$_6$H$_9$_α(T)+TRANS_HONO | 0 | 27.3 | -8.9 | -4.4 |
| R22 | C$_6$H$_{10}$+NO$_2$→C$_6$H$_9$_α(T)+ HNO$_2$ | 0 | 16.9 | 0.1 | 4.6 |
| R23 | C$_7$H$_{10}$+NO$_2$→C$_7$H$_9$_α(T)+TRANS_HONO | 0 | 26.9 | -10.9 | -7.1 |
| R24 | C$_7$H$_{10}$+NO$_2$→C$_7$H$_9$_α(T)+HNO$_2$ | 0 | 16.4 | -2.7 | 1.9 |

Table 1 presents a detailed analysis of the reactive energies, hence energy barriers, associated with H-atom abstractions from alkynes and dienes that are determined from the PES. The relative energies are calculated as the absolute energies subtracting the energies of the respective reactants. With this treatment, the relative energy of the TS indicates the energy barrier for that specific H-atom abstraction reaction. For alkynes, energy barrier for reactions yielding the same HNO$_2$ product generally follow a similar trend as that observed with the BDEs in Fig. 1, following the order of β(P) > α(P) > α(S) > α(T). The energy barrier for the reactions forming TRANS_HONO is consistently higher than the other two counterparts. In dienes, the potential energy at the ν site for forming transition states is significantly higher than those at the α site. Specifically, the energies required to form TRANS_HONO, HNO$_2$, and CIS_HONO at the ν sites (i.e., from C$_4$H$_6$) are 35.3 kcal/mol, 30.7 kcal/mol, and 29.7 kcal/mol, respectively. In contrast, at the α site, the energies (averaged from $C_5H_8$, $C_6H_{10}$ and $C_7H_{10}$) for forming TRANS_HONO, HNO$_2$, and CIS_HONO are approximately 28 kcal/mol, 19 kcal/mol, and 22 kcal/mol, respectively. Additionally, by comparing the energy barriers between $C_6H_{10}$ and $C_7H_{10}$, it can be seen that the energy barriers at the same site are within 0.5 kcal/mol (e.g., approximately 27.0 kcal/mol and 16.7 kcal/mol at the α(T) cite for forming TRANS_HONO and HNO$_2$, respectively)

### 3.3. Rate constants and rate rules

The rate coefficients for H-atom abstraction by NO$_2$, fitted into the Arrhenius expression, are summarized in Table 2. The rate coefficients forming different products at the same reaction site, in the temperature range of 298 – 2000 K, are illustrated in Figs. S3 – S5 in the Supplementary Material. In all the cases, the rate coefficient for the reaction forming TRANS_HONO is consistently smaller in the low temperature range, which is consistent with the energy barrier shown in Table 1. This difference decreases with increasing temperature.

Table 2 The rate coefficient for H-atom abstraction by NO$_2$ from alkynes and dienes to form the respective products and HNO$_2$ isomers (TRANS_HONO, HNO$_2$, CIS_HONO).

| No. | Reaction | A (cm³/mol*s) | n | Ea (cal/mol) |
|---|---|---|---|---|
| | Alkynes + NO$_2$ | | | |
| R1 | C$_3$H$_4$+NO$_2$→C$_3$H$_3$_α(P)+TRANS_HONO | 1.000E+00 | 3.707 | 28054.10 |
| R2 | C$_3$H$_4$+NO$_2$→C$_3$H$_3$_α(P)+HNO$_2$ | 1.000E+00 | 3.795 | 23170.45 |
| R3 | C$_3$H$_4$+NO$_2$→C$_3$H$_3$_α(P)+CIS_HONO | 1.000E+00 | 3.795 | 21896.31 |
| R4 | C$_4$H$_6$_1+NO$_2$→C$_4$H$_5$_β(P)+TRANS_HONO | 1.001E+00 | 3.933 | 30464.77 |
| R5 | C$_4$H$_6$_1+NO$_2$→C$_4$H$_5$_β(P)+HNO$_2$ | 1.001E+00 | 3.795 | 28695.59 |
| R6 | C$_4$H$_6$_1+NO$_2$→C$_4$H$_5$_β(P)+CIS_HONO | 1.001E+00 | 3.964 | 25809.74 |
| R7 | C$_4$H$_6$_1+NO$_2$→C$_4$H$_5$_α(S)+HNO$_2$ | 1.001E+00 | 3.603 | 19932.70 |
| R8 | C$_4$H$_6$_1+NO$_2$→C$_4$H$_5$_α(S)+CIS_HONO | 1.001E+00 | 3.689 | 19811.52 |
| R9 | C$_5$H$_8$+NO$_2$→C$_5$H$_7$_α(T)+TRANS_HONO | 1.001E+00 | 3.650 | 25086.67 |
| R10 | $C_5H_8+NO_2 \rightarrow C_5H_7\_\alpha(T)+HNO_2$ | 1.001E+00 | 3.640 | 17215.96 |
| R11 | $C_5H_8+NO_2 \rightarrow C_5H_7\_\alpha(T)+CIS\_HONO$ | 1.001E+00 | 3.926 | 18231.40 |
| Dienes + $NO_2$ | | | | |
| R12 | $C_4H_6\_13+NO_2 \rightarrow C_4H_5\_v(T)+TRANS\_HONO$ | 1.001E+00 | 3.938 | 32565.68 |
| R13 | $C_4H_6\_13+NO_2 \rightarrow C_4H_5\_v(T)+HNO_2$ | 1.002E+00 | 3.933 | 27257.57 |
| R14 | $C_4H_6\_13+NO_2 \rightarrow C_4H_5\_v(T)+CIS\_HONO$ | 1.000E+00 | 4.030 | 25063.31 |
| R15 | $C_5H_8\_13+NO_2 \rightarrow C_5H_7\_\alpha(P)+TRANS\_HONO$ | 1.001E+00 | 3.317 | 23831.78 |
| R16 | $C_5H_8\_13+NO_2 \rightarrow C_5H_7\_\alpha(P)+HNO_2$ | 1.001E+00 | 3.453 | 17667.69 |
| R17 | $C_5H_8\_13+NO_2 \rightarrow C_5H_7\_\alpha(P)+CIS\_HONO$ | 1.001E+00 | 3.529 | 18625.16 |
| R18 | $C_5H_8\_14+NO_2 \rightarrow C_5H_7\_\alpha(S)+TRANS\_HONO$ | 1.001E+00 | 3.383 | 24153.35 |
| R19 | $C_5H_8\_14+NO_2 \rightarrow C_5H_7\_\alpha(S)+HNO_2$ | 1.001E+00 | 3.641 | 15693.22 |
| R20 | $C_5H_8\_14+NO_2 \rightarrow C_5H_7\_\alpha(S)+CIS\_HONO$ | 1.000E+00 | 3.466 | 17517.12 |
| R21 | $C_6H_{10}+NO_2 \rightarrow C_6H_9\_\alpha(T)+TRANS\_HONO$ | 1.001E+00 | 3.257 | 22743.76 |
| R22 | $C_6H_{10}+NO_2 \rightarrow C_6H_9\_\alpha(T)+HNO_2$ | 1.001E+00 | 3.497 | 12451.33 |
| R23 | $C_7H_{10}+NO_2 \rightarrow C_7H_9\_\alpha(T)+TRANS\_HONO$ | 1.001E+00 | 3.250 | 22241.79 |
| R24 | $C_7H_{10}+NO_2 \rightarrow C_7H_9\_\alpha(T)+HNO_2$ | 1.001E+00 | 3.423 | 11807.62 |

Figure 3 illustrates the rate coefficients for H-atom abstraction by $NO_2$ from alkynes forming different products. For the reactions forming TRANS_HONO (Fig.3a), the difference in rate coefficients among all species is within four orders of magnitude at 298K, which narrows down to approximately one magnitude order at 2000K. The rate coefficient at the P sites is lower than that at the T site, regardless of the proximity to the triple bond. On the other hand, as depicted in Figs. 3b and 3c, the rate coefficients at low temperatures for reactions forming both $HNO_2$ and CIS_HONO show substantial differences between different reaction sites. The highest rate coefficients are observed for the reaction of $C_5H_8$ forming $C_5H_7\_\alpha(T)$, followed in descending order by $C_3H_4$ to $C_3H_3\_\alpha(P)$, $C_4H_6\_1$ to $C_4H_5\_\alpha(S)$, and lastly, $C_4H_6\_1$ to $C_4H_5\_\beta(P)$. Notably, the rate coefficients for $C_4H_6\_1$ at the $\beta(P)$ site are significantly lower than those at the $\alpha(P)$ site, which differs from the trends observed in Fig. 3a. Furthermore, the rate coefficient between $C_4H_6\_1$ forming $C_4H_5\_\alpha(S)$ and $C_3H_4$ forming $C_3H_3\_\alpha(P)$ differs by approximately two orders of magnitude, which quickly diminishes as temperature increases.

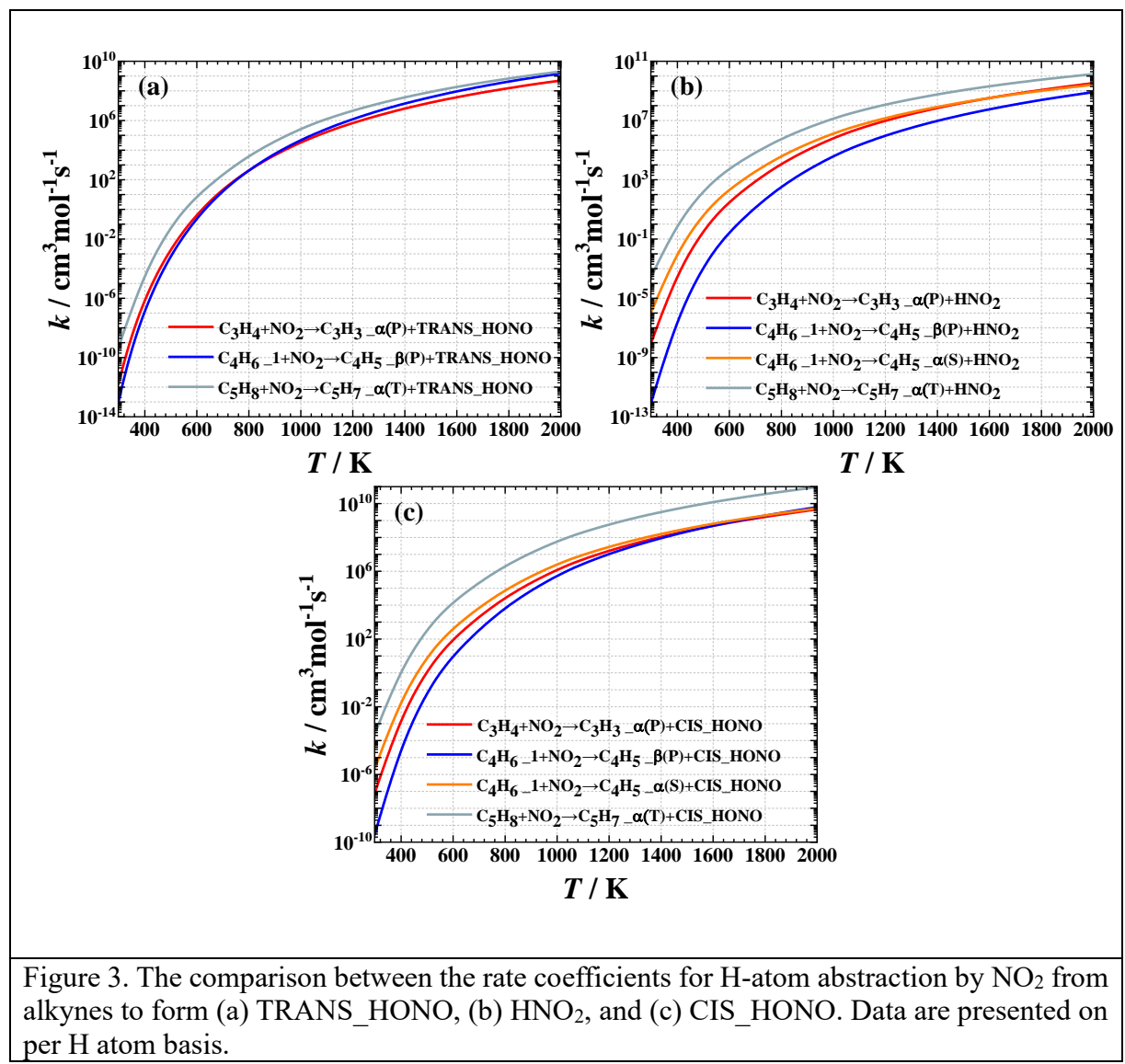

Figure 3. The comparison between the rate coefficients for H-atom abstraction by $NO_2$ from alkynes to form (a) TRANS_HONO, (b) $HNO_2$, and (c) CIS_HONO. Data are presented on per H atom basis.

Figure 4 focuses on comparing the rate coefficients for H-atom abstraction by the $NO_2$ from dienes. It is obvious from Figs. 4a and 4b that the rate coefficients for $C_6H_{10}$ forming $C_6H_9\_\alpha(T)$ closely mirror those for $C_7H_{10}$ forming $C_7H_9\_\alpha(T)$ across the temperature range studied, highlighting the consistent rate rules at the same reaction site among species of different sizes. At low temperatures, the rate coefficients are ranked from highest to lowest as follows: α(T) > α(S) > α(P) > v(T). This sequence highlights the significant influence of reaction site on the rate of H-atom abstraction. Furthermore, as can be seen in Fig. 4a, the rate coefficient for $C_5H_8\_13$ forming $C_5H_7\_\alpha(P)$ is slightly lower than that for $C_5H_8\_14$ forming $C_5H_7\_\alpha(S)$, primarily due to a small difference in energy barriers between these two reactions, which is around 0.2 kcal/mol (c.f. Table 1). As temperature increases, the difference in rate coefficients across all reactions diminishes. Nevertheless, it is worth noting that the rate coefficient for $C_4H_6\_13$ forming $C_4H_5\_v(T)$ dramatically increases with temperature, ultimately becoming the highest or nearly the highest among all reactions at 2000K. This could be due to the enhanced influences of hindered rotor effects at higher temperatures. Comparison between the rate coefficients for H-atom abstraction by $NO_2$ radical from different sites on the same molecule has also been conducted for $C_4H_6\_1$ (as can be seen in Fig. S5), where the rate coefficient at the $\beta(P)$ site is always several magnitudes lower than that at the $\alpha(S)$ site.

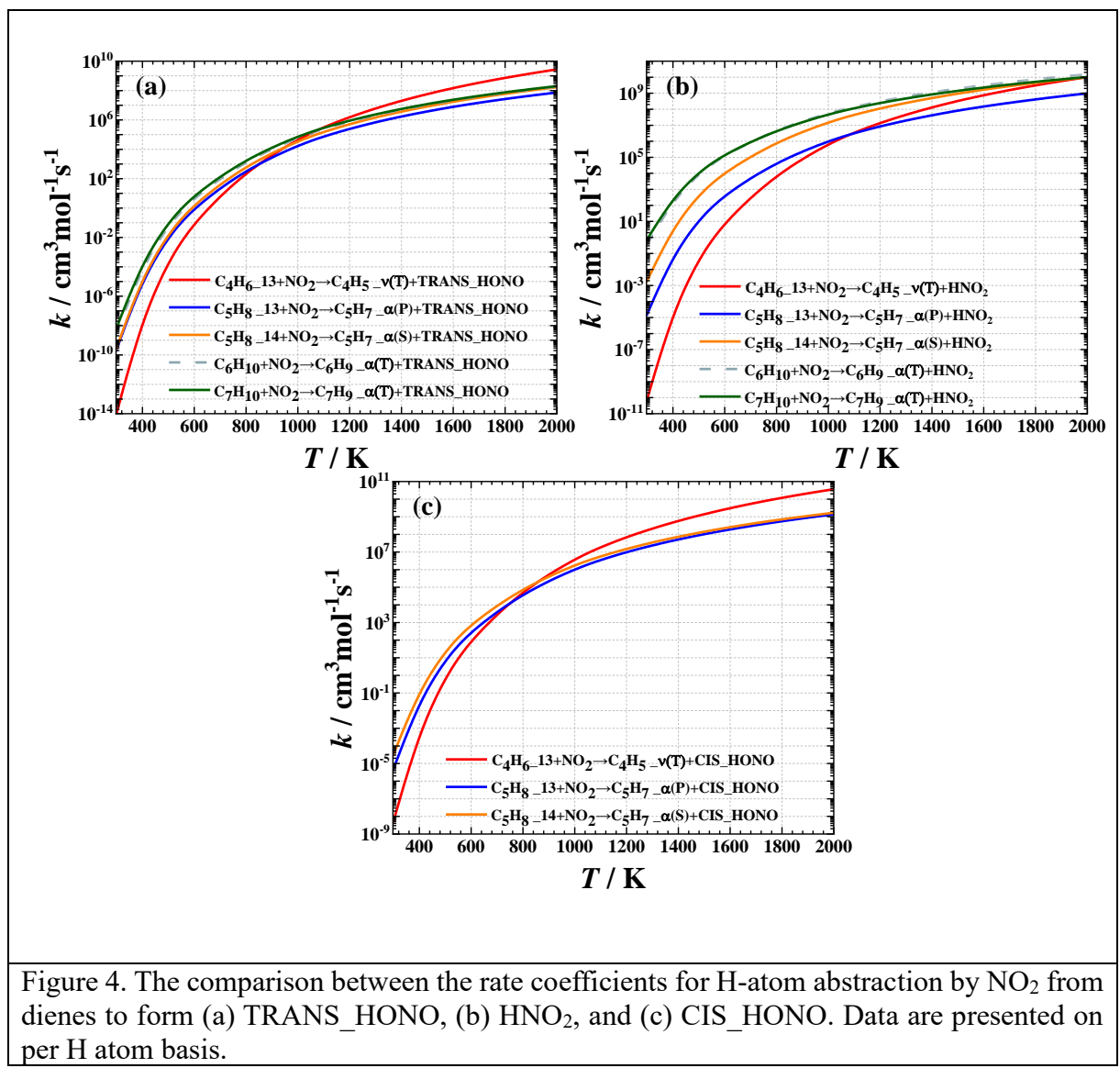

Figure 4. The comparison between the rate coefficients for H-atom abstraction by $NO_2$ from dienes to form (a) TRANS_HONO, (b) $HNO_2$, and (c) CIS_HONO. Data are presented on per H atom basis.

### 3.4. Rate constants and branching ratios

Figure 6 illustrates the branching ratios for H-atom abstraction by the $NO_2$ from alkynes to form $HNO_2$ isomers. Immediately seen in Fig. 6 is the dominancy of the CIS_HONO channel, which exhibits a branching ratio that is considerably higher than the other two channels. This is most obvious for $C_4H_6$ (Fig. 6b) where the branching ratio for the CIS_HONO channel reaches nearly 1 at below 400 K. Differences are also observed between the three alkynes.

Specifically, the branching ratios for CIS_HONO at the α-P site of $C_3H_4$ and the β-P site of $C_4H_6$ decrease with rising temperatures, as seen in Figs. 6a and 6b, respectively, whereas that at the α-T site of $C_5H_8$ increases with temperature. Consequently, the dependence of branching ratio on temperature for the $HNO_2$ channel also displays opposite trends between the α-T site $C_5H_8$ and the other two sites of $C_3H_4$ and $C_4H_6$. Notably, for $HNO_2$, the branching ratio increases significantly for $C_3H_4$ at the α-P site from about 10% to 40% as the temperature rises to 2000 K, while for $C_4H_6$ at the β-P site, it falls below that of TRANS_HONO, stabilizing at around 10%. Across all species, the proportions of TRANS_HONO remain relatively low, typically not exceeding 20%, indicating that this pathway is less favored compared to the pathways forming CIS_HONO and $HNO_2$.

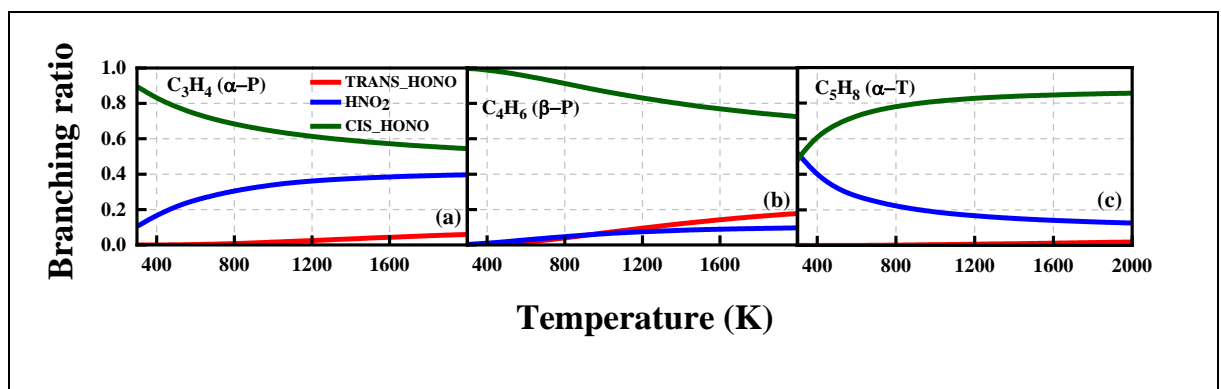

Figure 6. The comparison between the branching ratio for H-atom abstraction by $NO_2$ from alkynes to form $HNO_2$ isomers (TRANS_HONO, $HNO_2$, CIS_HONO), (a) $C_3H_4$(α-P), (b) $C_4H_6$(β-P), (c) $C_5H_8$(α-T).

Figure 7 further shows the branching ratios for H-atom abstraction by the $NO_2$ from dienes. These branching ratios correlate closely with the corresponding energy barriers in Table 1. For instance, the branching ratios of CIS_HONO and $HNO_2$ for $C_5H_8\_13$ at the α-P site intersect at approximately 870K, as illustrated in Fig. 7b and the rate coefficients for these reactions also intersect at this temperature (see Fig. S4b). This intersection highlights a critical temperature where the preferential formation of the HNO$_2$ isomers shifts. The trends of branching ratio at the ν-T site of C$_4$H$_6$_13 are similar to those observed in Fig. 6, where the CIS_HONO channel is the most dominant. However, a closer observation of Fig. 7 highlights the considerably different branching ratios at the α site between alkynes (Figs. 6a and 6c) and dienes (Figs. 7b and 7c). Specifically, at the α site of dienes, the branching ratio is mostly favored toward the HNO$_2$ channel, rather than the CIS_HONO channel as observed for alkynes. The TRANS_HONO channel is consistently the least important for both alkynes and dienes.

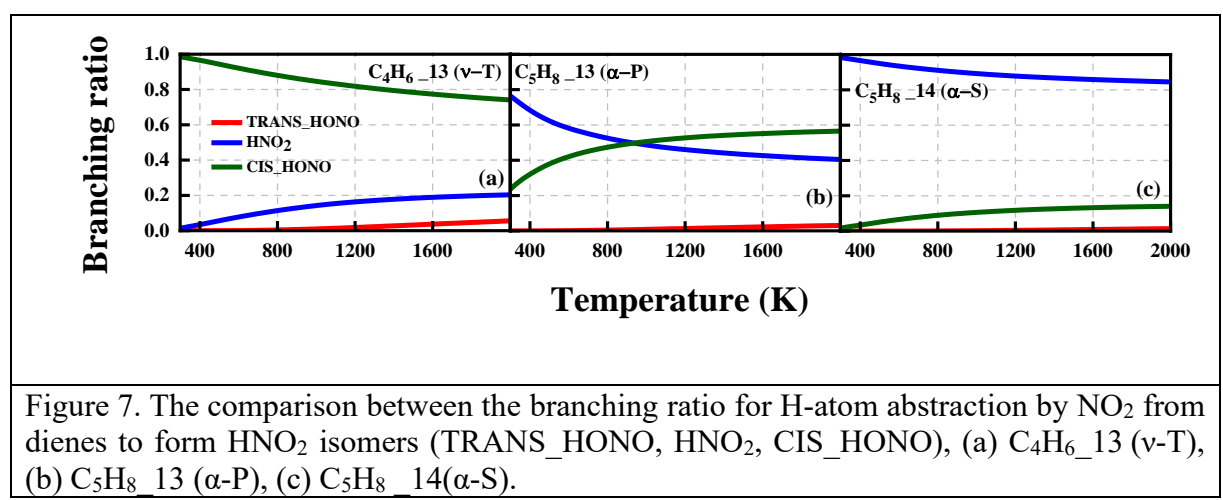

Figure 7. The comparison between the branching ratio for H-atom abstraction by NO$_2$ from dienes to form HNO$_2$ isomers (TRANS_HONO, HNO$_2$, CIS_HONO), (a) C$_4$H$_6$_13 (ν-T), (b) C$_5$H$_8$_13 (α-P), (c) C$_5$H$_8$_14 (α-S).

Figure 8 illustrates the branching ratios of alkynes and dienes versus carbon number across various temperatures. At 600 K (Fig. 8a), the branching ratio for CIS_HONO starts at about 75% for alkynes with three carbon atoms, increases to approximately 97% for those with four carbon atoms, and then decreases slightly to about 70% for five carbon atoms. In contrast, the branching ratio for HNO$_2$ displays a reversed trend. These trends are consistent at different temperatures, as depicted in Figs. 8b – 8d. Greater shifts in branching ratio cross different carbon atom numbers are observed for dienes (Fig. 8 – right panel). At all temperatures studied, the branching ratio of CIS_HONO is significantly higher than that of $HNO_2$ for dienes with four carbon atoms. However, when the carbon count increases to five, the trend reverses, with the branching ratio of CIS_HONO falling below that of $HNO_2$.

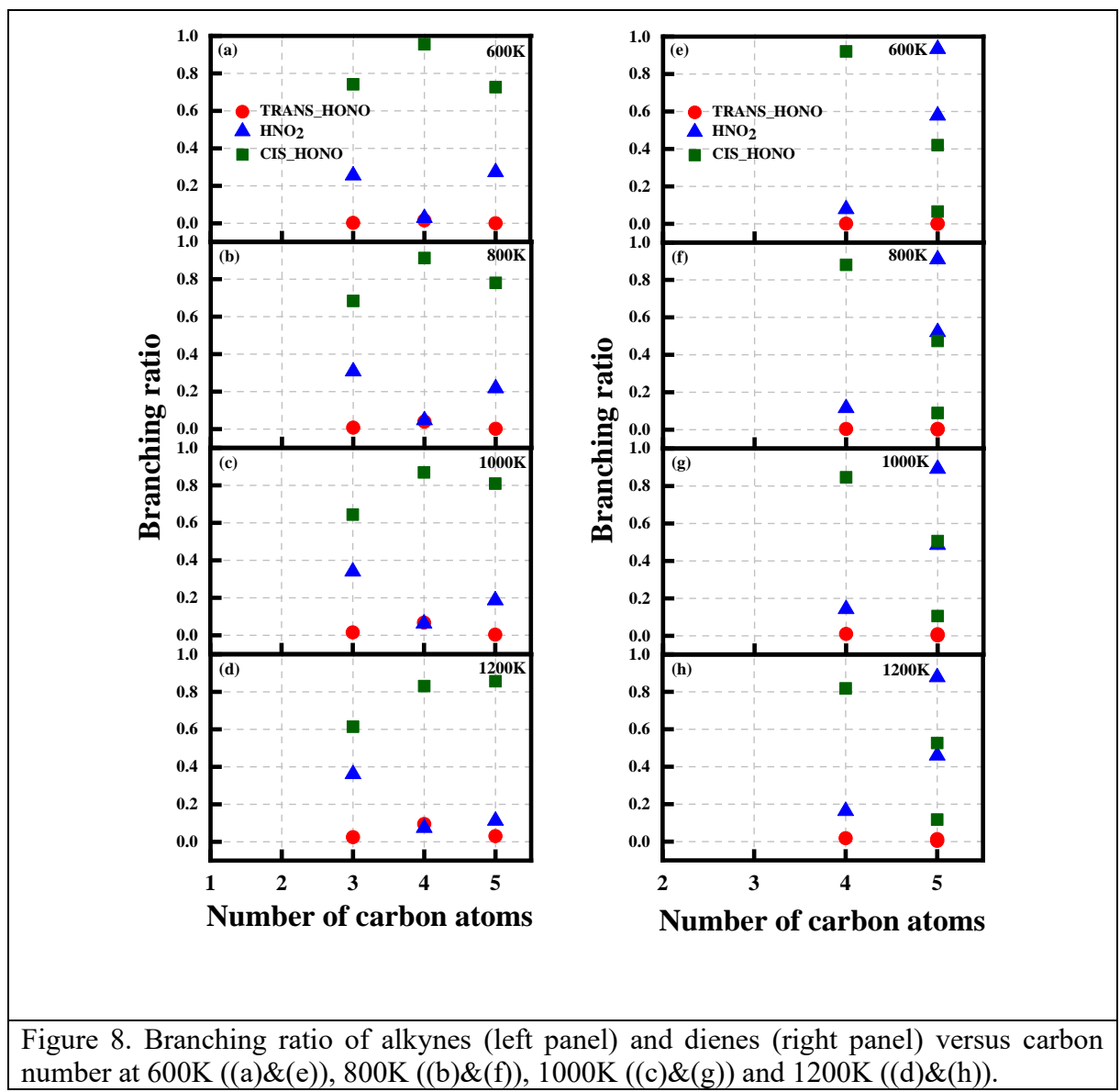

Figure 8. Branching ratio of alkynes (left panel) and dienes (right panel) versus carbon number at 600K ((a)&(e)), 800K ((b)&(f)), 1000K ((c)&(g)) and 1200K ((d)&(h)).

## 3.5. Model implementations and implications

To investigate the influences of the studied reactions on model predictions, the calculated rate parameters are incorporated individually for each species into the detailed gasoline surrogate chemistry model [37]. The model without and with the updated rate parameters is referred to as "original model" and "updated model" in the following. It should be noted that the purpose of model implementations is to illustrate the impact of the calculated reactions, rather than to propose a new model. The experimental conditions from [14] are adopted, where a rapid compression machine was used to study the interactions between $NO_x$ and unsaturated hydrocarbons with strong $NO_x$ blending effects on ignition propensity observed.

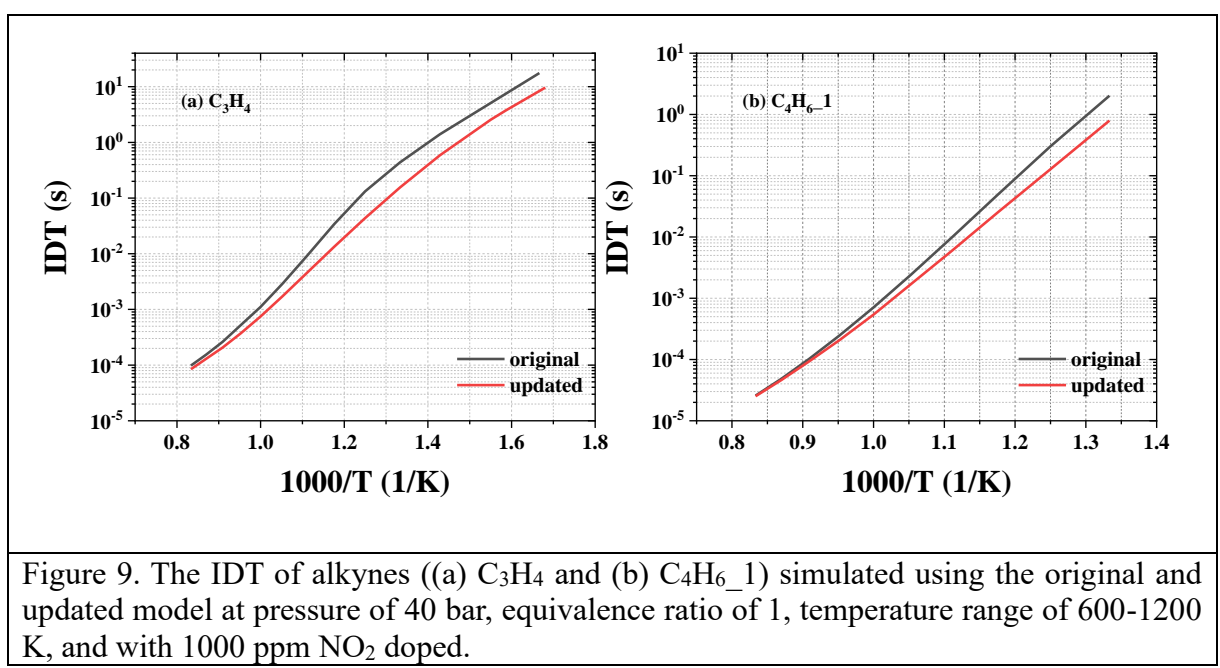

Figure 9. The IDT of alkynes ((a) $C_3H_4$ and (b) $C_4H_6\_1$) simulated using the original and updated model at pressure of 40 bar, equivalence ratio of 1, temperature range of 600-1200 K, and with 1000 ppm $NO_2$ doped.

Figure 9 compares the simulated ignition delay times (IDT) between the original and updated models for alkynes. It can be seen that the original model and the updated model show significant differences with greater differences observed at low temperatures, indicating the critical role of the studied reactions in determining model reactivity. Specifically, the updated model becomes more reactive than the original model, advancing the ignition delay time by up to 0.5 magnitude order. Besides, the negative temperature coefficient (NTC) behavior is seemingly weakened in the updated model for $C_3H_4$, as shown in Fig. 9a.

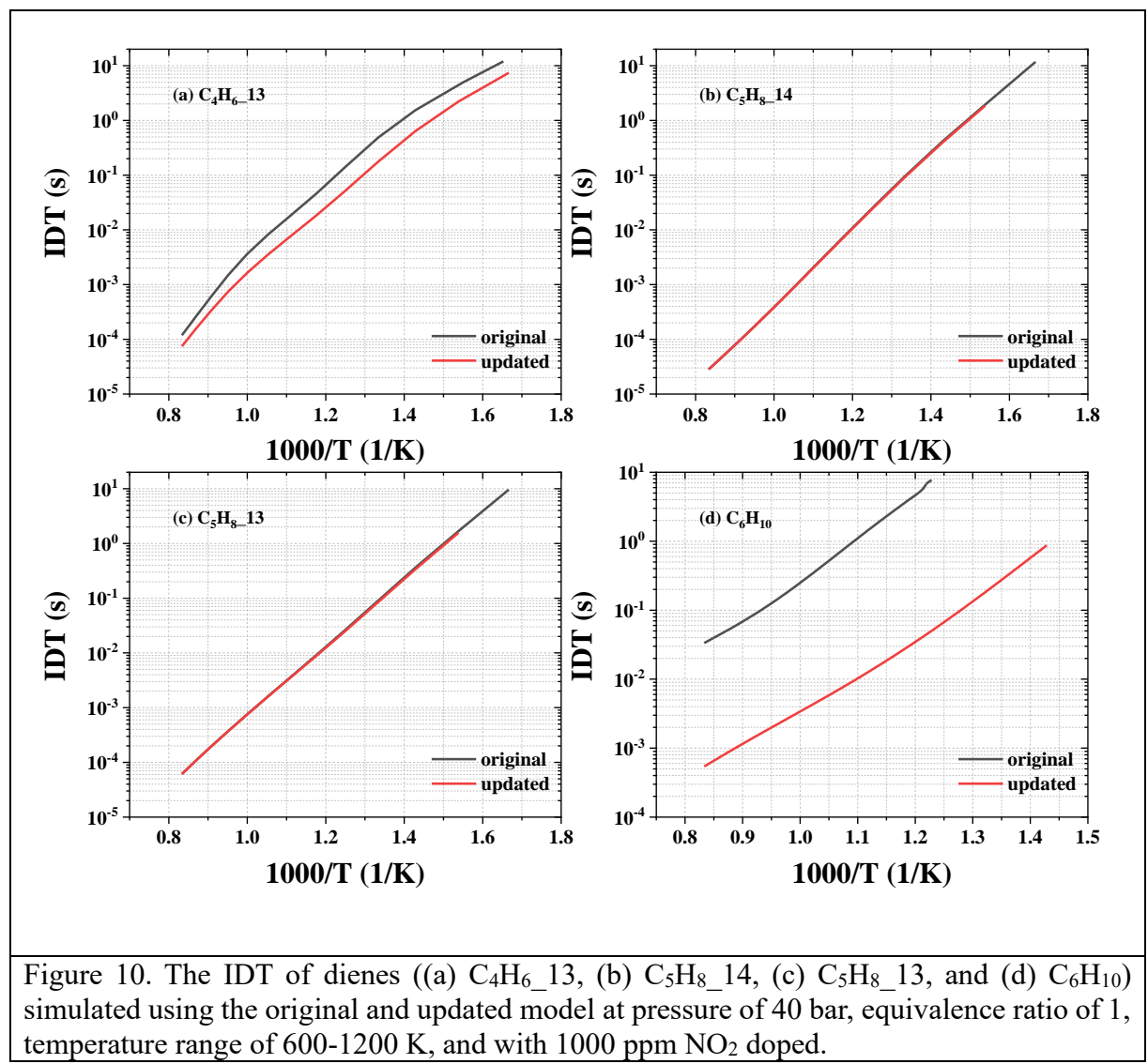

Figure 10. The IDT of dienes ((a) $C_4H_6\_13$, (b) $C_5H_8\_14$, (c) $C_5H_8\_13$, and (d) $C_6H_{10}$) simulated using the original and updated model at pressure of 40 bar, equivalence ratio of 1, temperature range of 600-1200 K, and with 1000 ppm $NO_2$ doped.

Figure 10 shows the comparison of the simulated IDT for dienes between the original and updated models. It is clear from Fig. 10 that the updated rate parameters greatly influence the model prediction results for dienes, highlighting the need to accurately represent these reactions in chemistry models. Similar to the results in Fig. 9, with incorporating the updated rate parameters, the model becomes more reactive, with more pronounced change in model predictions observed for $C_4H_6\_13$ and $C_6H_{10}$ than for $C_5H_8\_14$ and $C_5H_8\_13$. The change in model reactivity is surprisingly large for $C_6H_{10}$ (e.g., IDT advanced by ~2 magnitude orders, as shown in Fig. 10d), indicating that its chemistry was ill-conditioned in the original model and requires further updates.

To investigate the underlying kinetics governing the model reactivity after incorporating the updated rate parameters, sensitivity analysis is further conducted at the same conditions as in Figs. 9 and 10, but only at two representative temperatures, i.e., 700 and 1100 K. Two representative species, namely $C_4H_6\_1$ for alkynes and $C_4H_6\_13$ for dienes, are selected. The sensitivity analysis coefficient is defined as $S_{rel} = \ln(\frac{\tau^\Delta}{\tau})/\ln(\frac{k^\Delta}{k})$, where $\tau^\Delta$ is the gas temperature (over the whole ignition delay period) after multiplying the original rate constant by 2, i.e., $k^\Delta = 2*k$, and $\tau$ is the original mixture temperature. Negative sensitivity coefficient indicates an inhibiting effect on mixture reactivity, while the positive sensitivity coefficient indicates a promoting effect. The computed sensitivity coefficients for the 16 most sensitive reactions are presented for each mixture at each condition.

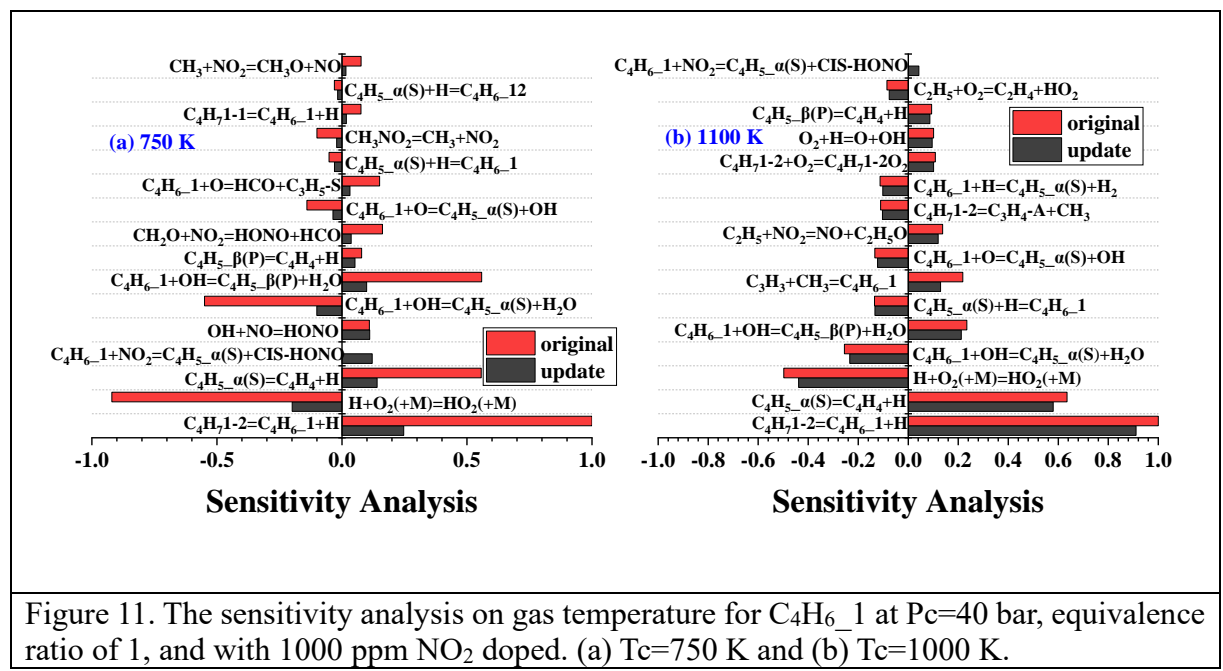

Figure 11. The sensitivity analysis on gas temperature for $C_4H_6\_1$ at Pc=40 bar, equivalence ratio of 1, and with 1000 ppm $NO_2$ doped. (a) Tc=750 K and (b) Tc=1000 K.

As shown in Fig. 11a, at 750 K, the sensitivity coefficients computed by the original model are considerably greater than those computed by the updated model, whereas those at 1100 K are quite similar between the original and updated models. This is consistent with the IDT results in Fig. 9b, where the change in predicted IDT between the two models is almost the highest at 750 K while the lowest at 1100 K. Overall, both the promoting and inhibiting effects of the most sensitive reactions at 750 K are suppressed in the updated model, counteracting each other. However, the fifth most sensitive reaction, namely $C_4H_6\_1+NO_2=C_4H_5\_\alpha(S)+CIS\_HONO$, displays a strong promoting effect on autoignition reactivity in the updated model, whereas this reaction was not available in the original model. The produced CIS_HONO quickly decomposes to OH and NO, which further promotes ignition reactivity, as can be seen in Fig. 11a where OH+NO=HONO exhibits a positive sensitivity coefficient. The sensitivity coefficients for OH+NO=HONO remain similar between the updated and original models, which is different from the other reactions. This is primarily due to the increased contribution to HONO production via $C_4H_6\_1+NO_2=C_4H_5\_\alpha(S)+CIS\_HONO$. These contributions eventually lead to the increased autoignition reactivity with the updated model. The trends observed at 750 K for $C_4H_6\_1+NO_2=C_4H_5\_\alpha(S)+CIS\_HONO$ are also observed at 1100 K, but less pronounced, as can be inferred from Fig. 11b.

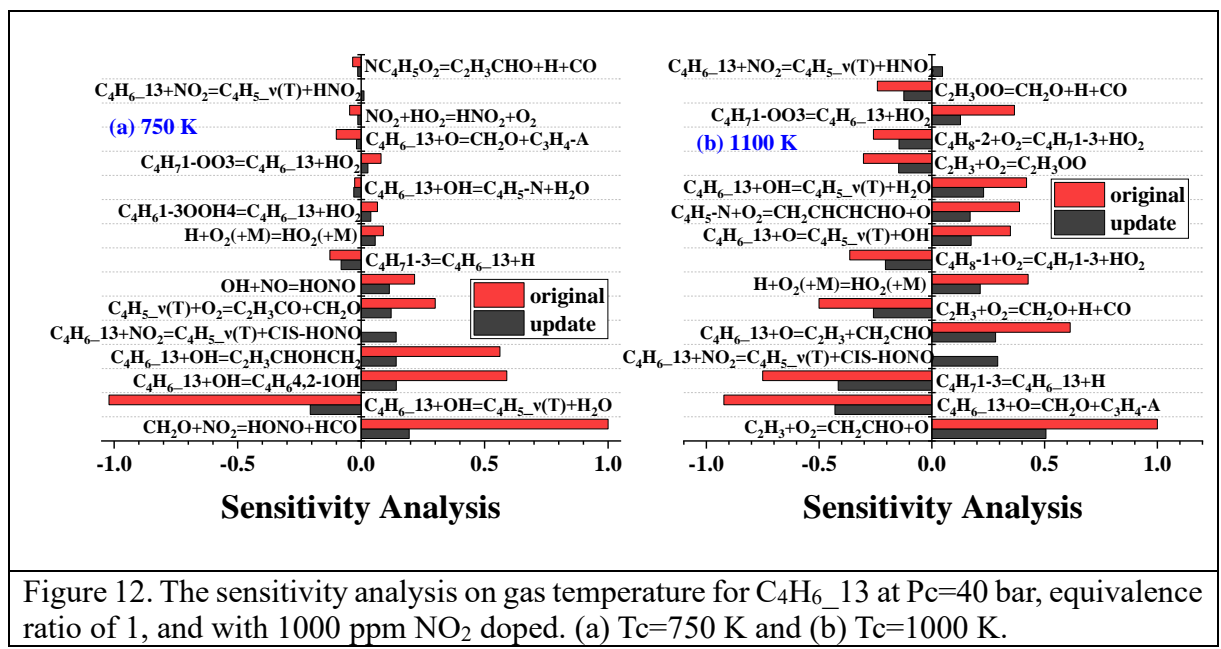

Figure 12. The sensitivity analysis on gas temperature for $C_4H_6\_13$ at Pc=40 bar, equivalence ratio of 1, and with 1000 ppm $NO_2$ doped. (a) Tc=750 K and (b) Tc=1000 K.

The sensitivity analysis results for $C_4H_6\_13$, as shown in Fig. 12, are somewhat different from those in Fig. 11. First, the difference in sensitivity coefficients between the updated and original models is significant at both 750 K and 1100 K, with relatively higher difference seen at 750 K. This, again, corroborates with the simulated IDT results (c.f., Fig. 10a), where the updated model shows different autoignition reactivity at both 750 K and 1100 K. Second, apart from the dominant promoting effect from the reaction forming CIS-HONO in the updated model, namely $C_4H_6\_13+NO_2=C_4H_5\_v(T)+CIS\_HONO$, there is also a relatively smaller promotion effect from $C_4H_6\_13+NO_2=C_4H_5\_v(T)+HNO_2$. The different levels of contribution from the CIS_HONO and $HNO_2$ channels are supported by the branching ratios in Fig. 7a where the CIS_HONO channel is consistently the most dominant pathway at all temperatures studied. The promotion effects from these two H-atom abstraction pathways by $NO_2$ explain the increased autoignition reactivity in the updated model.

Flux analyses are further conducted at the same conditions as sensitivity analysis at 1%

fuel consumption, which are summarized in Figs. 13 and 14 for $C_4H_6\_1$ (alkyne) and $C_4H_6\_13$ (diene). The percentages shown in Figs. 13 and 14 are computed as the ratio of the rate of consumption for that pathway to the total rate of consumption.

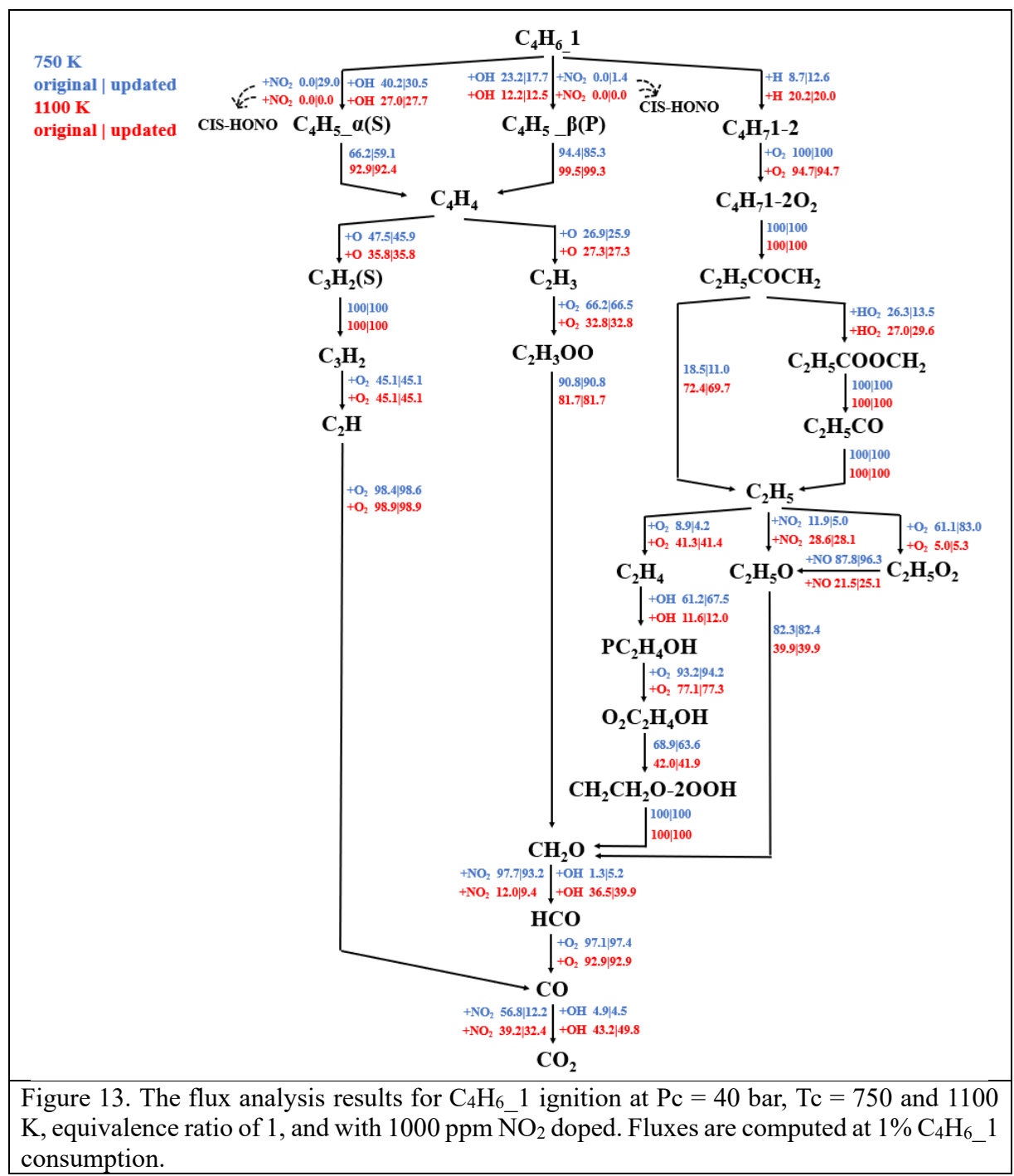

Figure 13. The flux analysis results for $C_4H_6\_1$ ignition at Pc = 40 bar, Tc = 750 and 1100 K, equivalence ratio of 1, and with 1000 ppm $NO_2$ doped. Fluxes are computed at 1% $C_4H_6\_1$ consumption.

As can be seen in Fig. 13, the H atom of $C_4H_6\_1$ is mainly abstracted by OH and H radicals in the original model, leading to the formation of $C_4H_5\_\alpha(S)$, $C_4H_5\_\beta(P)$ and $C_4H_71$-2. However, with the H-atom abstraction reactions by $NO_2$ incorporated into the updated model, over 30% of $C_4H_6\_1$ is consumed by $NO_2$ at 750 K, with 29% and 1.4% of $C_4H_6\_1$ reacted to form $C_4H_5\_\alpha(S)$ and $C_4H_5\_\beta(P)$, respectively. As a result, $C_4H_6\_1$ consumption by OH is reduced at this temperature, e.g., from 40.2% to 30.5% for the formation of $C_4H_5\_\alpha(S)$. As can be seen from Fig. 11a that $C_4H_6\_1+NO_2=C_4H_5\_\alpha(S)+CIS\_HONO$ is a promoting reaction, while $C_4H_6\_1+OH=C_4H_5\_\alpha(S)+H_2O$ is an inhibiting reaction. The increased flux in the former reaction and decrease in the later reaction will greatly promote ignition reactivity, as observed in Fig. 9a. The shifts in fluxes for the $C_4H_6\_1$ consuming pathways observed at 750 K are not observed at 1100 K, where the H-atom abstractions by $NO_2$ show no contributions. Instead, great changes are seen for the consumption of $CH_2O$ and CO, where the flux through the $NO_2$-involving channel is reduced in the updated model. Nevertheless, these two reactions are not identified as the most sensitive reactions at 1100 K, as shown in Fig. 11b.

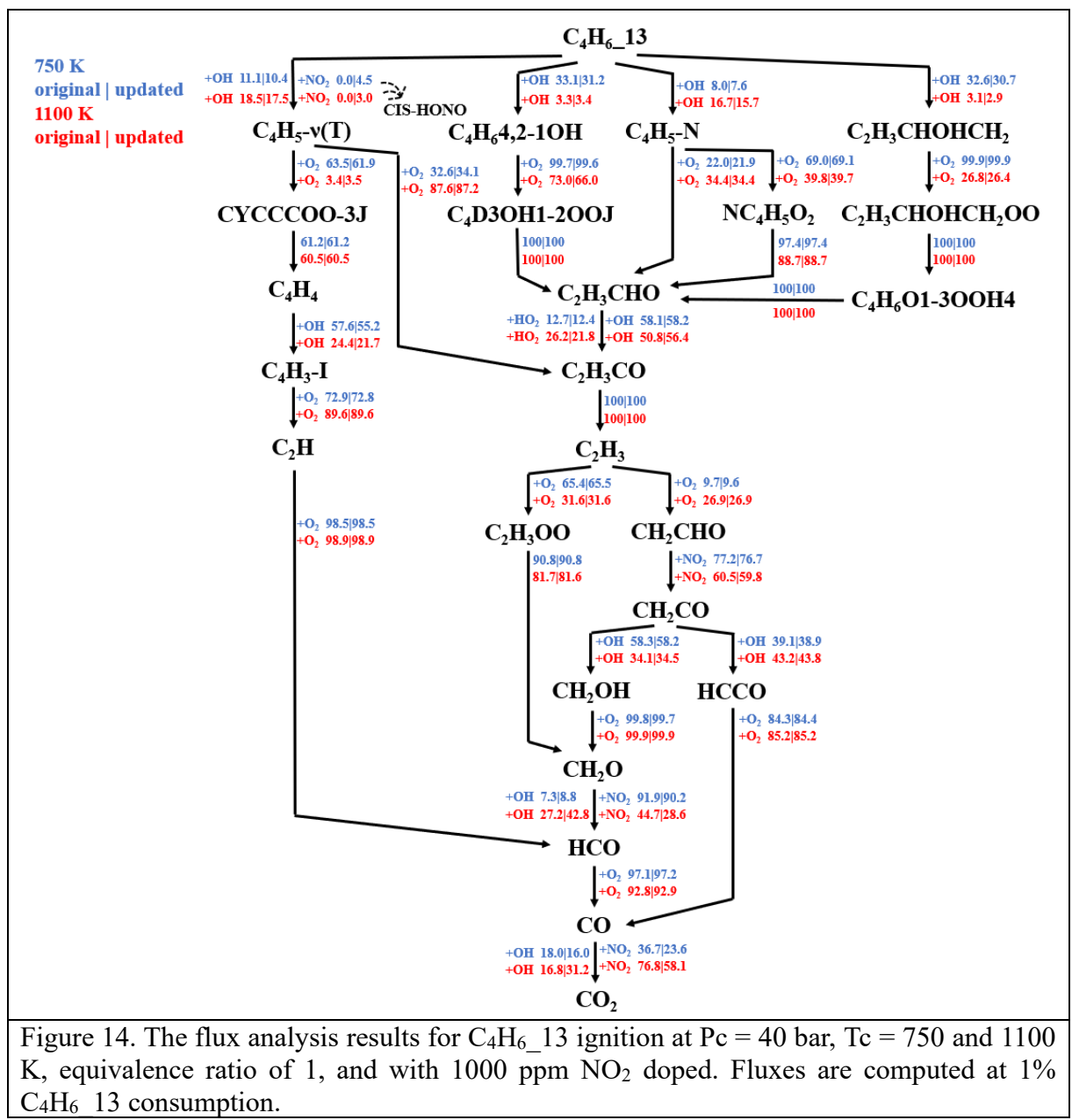

Figure 14. The flux analysis results for $C_4H_6\_13$ ignition at $P_c$ = 40 bar, $T_c$ = 750 and 1100 K, equivalence ratio of 1, and with 1000 ppm $NO_2$ doped. Fluxes are computed at 1% $C_4H_6\_13$ consumption.

Similar to the results in Fig. 13, Fig. 14 shows that the consuming pathways of $C_4H_6\_13$ shift toward H-atom abstraction by $NO_2$ forming $C_4H_5\_v(T)$ and away from the H-atom abstraction by OH in the updated model at 750 K. This leads to an enhanced promoting effect from $C_4H_6\_13+NO_2=C_4H_5\_v(T)+CIS\_HONO$ and a suppressed inhibiting effect from $C_4H_6\_13+OH=C_4H_5\_v(T)$ (or $C_4H_5\_N$) $+H_2O$, resulting in an increased model reactivity. Different from Fig. 13, the H-atom abstraction by $NO_2$ forming $C_4H_5\_v(T)$ demonstrates obvious contributions to $C_4H_6\_13$ consumption even at 1100 K. This explains the results in Fig. 12b, where $C_4H_6\_13+NO_2=C_4H_5\_v(T)+CIS\_HONO$ is still among the top promoting reactions at 1100 K.

## 4. Conclusions

This work presents a systematic study of H-atom abstractions by $NO_2$ from different sites of C3–C7 alkynes and dienes to form different $HNO_2$ isomers (i.e., TRANS_HONO, $HNO_2$, and CIS_HONO). The geometry optimizations and vibrational frequencies of all the involved species are conducted at M06–2X/6–311++G(d,p) level of theory. The SPEs are determined at DLPNO-CCSD(T)/cc-pVDZ level of theory. The energy barriers of the investigated 24 reactions are obtained. The temperature dependent rate coefficients of these 24 reactions are proposed at the temperature range of 298.15–2000 K, based on conventional transition state theory with unsymmetric Eckart tunneling corrections. Updated rate constants for these reactions are further incorporated into a recently updated chemistry model, where their influences on model performance are investigated via comprehensive kinetic modeling. Sensitivity and flux analyses are further conducted to explore the chemical kinetics governing changes in model performances. The primary conclusions from this study are:

- Energy barriers at the primary (P), secondary (S), and tertiary (T) sites of alkynes and dienes decrease in the order of P > S > T. The C=C and C≡C functional groups considerably lower the energy barriers for abstracting the adjacent hydrogen atoms at the α carbon site, while H-atom abstractions at the C=C sites exhibit significantly higher energy barriers than all other sites. For all the species and reaction cites studied, the energy barriers for producing TRANS_HONO are consistently higher than those forming $HNO_2$ and CIS_HONO.

- After incorporating the rate parameters of the calculated reactions, the kinetic model becomes considerably more reactive in predicting the autoignition of all alkynes and dienes, particularly for $C_6H_{10}$, with greater changes observed at low temperatures than at intermediate to high temperatures.

- Sensitivity and flux analyses reveal the kinetics governing the changes in model reactivity. With the updated rate parameters, there is an obvious increase and decrease in the contribution of H-atom abstractions by $NO_2$ and OH radicals, respectively, with the former and latter pathways being the top reactivity-promoting and reactivity-inhibiting reactions, which is mainly responsible for the increased model reactivity.

- This study highlights the critical role of H-atom abstractions from alkynes and dienes by $NO_2$, which have been missing from existing chemistry models. The obtained results emphasize the need for adequately representing these kinetics in new alkyne and diene chemistry models to be developed, by using the rate parameters determined in this study. There is also an urgent need for future experimental studies of fundamental combustion properties for $NO_2$-doped alkyne and diene mixtures.

**Acknowledgments**

This material is based on work supported by the Research Grants Council of Hong Kong Special Administrative Region, China, under PolyU P0046985 for ECS project funded in 2023/24 Exercise and P0050998, and by the Natural Science Foundation of Guangdong Province under 2023A1515010976 and 2024A1515011486.

## Declaration of Competing Interests

The authors declare no competing interests.

## Author Contributions

Z.G., H.W., R.T. and S.C. conceived and designed the study and performed the calculations, with technical inputs from X.R., M.W., T.Z., G.L. and H.G. S.C. supervised all work and acquired funding supports for this study. All authors contributed to writing the paper.